%% file: sample-sigconf.tex
\documentclass[sigconf]{acmart}
\usepackage[linesnumbered,ruled,vlined]{algorithm2e}
\usepackage{multirow}
\usepackage{dsfont}
\usepackage{url}
\usepackage{helvet}  
\usepackage{courier}  
\usepackage{makecell}
\usepackage{subcaption}
\usepackage{multirow}
\usepackage{todonotes}
\usepackage{enumitem}
\usepackage{makecell}
\usepackage{subcaption}
\usepackage[utf8]{inputenc} 
\usepackage{url}            
\usepackage{booktabs}       
\usepackage{amsfonts}       
\usepackage{nicefrac}       
\usepackage{microtype}      
\usepackage{xcolor}         
\usepackage{dsfont} 
\DeclareCaptionStyle{ruled}{labelfont=normalfont,labelsep=colon,strut=off} 
\usepackage{algorithmic}

\usepackage{amssymb}

\usepackage{newfloat}
\usepackage{colortbl}
\definecolor{myyellow}{rgb}{1,1, 0.6}
\definecolor{myorange}{rgb}{1, 0.8, 0.6}
\definecolor{myred}{rgb}{1, 0.6, 0.6}
\AtBeginDocument{%
  \providecommand\BibTeX{{%
    \normalfont B\kern-0.5em{\scshape i\kern-0.25em b}\kern-0.8em\TeX}}}

\setcopyright{acmcopyright}
\copyrightyear{2023}
\acmYear{2023}
\setcopyright{acmlicensed}\acmConference[WWW '23]{Proceedings of the ACM Web Conference 2023}{May 1--5, 2023}{Austin, TX, USA}
\acmBooktitle{Proceedings of the ACM Web Conference 2023 (WWW '23), May 1--5, 2023, Austin, TX, USA}
\acmPrice{15.00}
\acmDOI{10.1145/3543507.3583451}
\acmISBN{978-1-4503-9416-1/23/04}

\acmConference[WWW '23]{TheWebConf}{April 30 -- May 4,
  2023}{Austin, Texas, USA}
\acmBooktitle{WWW '23: ACM Symposium on Neural Gaze Detection,
  April 30 -- May 4, 2023, USA}
\acmISBN{978-1-4503-XXXX-X/23/02}

\usepackage{subcaption}
\usepackage{multirow}
\usepackage{todonotes}
\usepackage{enumitem}

\usepackage{bbding}

\begin{document}

\title{DUET: A Tuning-Free Device-Cloud Collaborative Parameters \\ Generation Framework for Efficient Device Model Generalization}

\author{Zheqi Lv$^{1\dagger}$, Wenqiao Zhang$^{2\dagger}$, Shengyu Zhang$^1$, Kun Kuang$^{1,3*}$, Feng Wang$^4$, Yongwei Wang$^5$, Zhengyu Chen$^1$, Tao Shen$^1$, Hongxia Yang$^4$, Beng Chin Ooi$^2$, Fei Wu$^{1,5,6*}$}
\thanks{$\dagger$These authors contributed equally to this research.\\ *Corresponding authors.}
\affiliation{
  \institution{
\textsuperscript{$1$}Zhejiang University,
\textsuperscript{$3$}Key Laboratory for Corneal Diseases Research of Zhejiang Province, \textsuperscript{$4$}Alibaba Group \country{China} \\
\textsuperscript{$2$} National University of Singapore \country{Singapore} \\
\textsuperscript{$5$}Shanghai Institute for Advanced Study of Zhejiang University, \textsuperscript{$6$}Shanghai AI Laboratory \country{China} \\
}
}
\email{
{zheqilv,sy_zhang,kunkuang,chenzhengyu,tao.shen,wufei}@zju.edu.cn,wenqiao@nus.edu.sg
}
\email{{windpls,firewater1984}@gmail.com,yongwei.wang@ntu.edu.sg,ooibc@comp.nus.edu.sg}

\renewcommand{\shortauthors}{Zheqi Lv et al.}



\input{tex/0abstract}
\maketitle
\input{tex/1introduction}
\input{tex/2related_works}
\input{tex/3method}

\input{tex/4experiment}
\input{tex/5conclusion}

\input{tex/acknowledgement}
\bibliographystyle{ACM-Reference-Format}
\bibliography{sample-base}

\end{document}

%% file: tex/0abstract.tex
\begin{abstract}
\label{sec:abstract}
\begin{sloppypar}
Device Model Generalization (DMG) is a practical yet under-investigated research topic for on-device machine learning applications. It aims to improve the generalization ability of pre-trained models when deployed on resource-constrained devices, such as improving the performance of pre-trained cloud models on smart mobiles. While quite a lot of works have investigated the \emph{data distribution shift} across clouds and devices, most of them focus on model fine-tuning on personalized data for individual devices to facilitate DMG. 
Despite their promising, these approaches require on-device re-training, which is practically infeasible due to the overfitting problem and 
high time delay when performing gradient calculation on real-time data.
In this paper, we argue that the computational cost brought by fine-tuning can be rather unnecessary. 
We consequently present a novel perspective to improving DMG without increasing computational cost, \textit{i.e.}, device-specific parameter generation which directly maps data distribution to parameters.
Specifically, we propose an efficient \textbf{\underline{D}}evice-clo\textbf{\underline{U}}d collaborative paramet\textbf{\underline{E}}rs genera\textbf{\underline{T}}ion  framework (\textbf{DUET}). 
DUET is deployed on a powerful cloud server that only requires the low cost of forwarding propagation and low time delay of data transmission between the device and the cloud. By doing so, DUET can rehearse the device-specific model weight realizations conditioned on the personalized real-time data for an individual device. Importantly, our DUET elegantly connects the cloud and device as a ``duet'' collaboration, frees the DMG from fine-tuning, and enables a faster and more accurate DMG paradigm.
We conduct an extensive experimental study of DUET on three public datasets, and the experimental results confirm our framework’s effectiveness and generalisability for different DMG tasks.
\end{sloppypar}
\end{abstract}

%% file: tex/1introduction.tex
\section{Introduction}
\label{sec:introduction}

\begin{figure*}[t]
\includegraphics[width=1\textwidth]{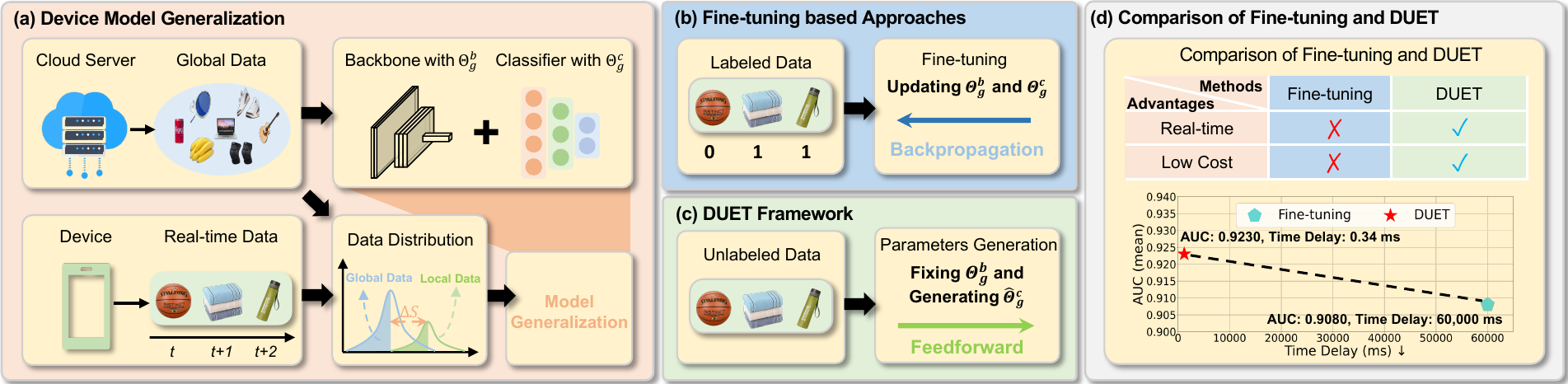}
\vspace{-0.5cm}
\centering\caption{ (a) describes the device model generalization in device-cloud collaboration, $\Delta S$ indicates the data distribution shift of global and local data. (b) and (c) are overviews of fine-tuning based approaches and our DUET, respectively. $\Omega_b$ and $\Omega_c$ respectively denoted the parameters of backbone and classifier. (d) is the comparison of fine-tuning and DUET (\textbf{\texttt{Time Delay: 0.34ms (DUET) $\ll$ 60,000ms (Fine-tuning)), AUC: 0.9230 (DUET) $>$ 0.9080 (Fine-tuning))}}. }
\label{fig1}
\vspace{-0.3cm}
\end{figure*}

The high performance of Deep Neural Networks (DNNs)~\cite{ref:vggnet,ref:resnet} is tempered by the huge parameter size of intricate design patterns and the demand for high computational costs. This situation greatly hinders the application of intelligent services in mobile phones and Internet of Things (IoT) devices since their hardware resources are tightly constrained by the form factor, battery, and heat dissipation. 
Therefore, on-device machine learning (DML) that goes beyond DNNs by exploiting lightweight neural networks (LNNs) for task-specific learning and inference on devices is gaining traction, such as MobileNets~\cite{ref:mobilenet,ref:mobilenetv2,ref:mobilenetv3}, DIN~\cite{ref:din}, SASRec~\cite{ref:sasrec}, GRU4Rec~\cite{ref:gru4rec}. 
Along with the rapid development of cloud computing, the predominant DML paradigm is not simply learning on the device. It often collaboratively connects with a powerful cloud server with perceivably limitless resources. 
As shown in Figure~\ref{fig1}(a), the cloud trains a global primary model conditioned on rich data collected from different devices, and then the local device performs inference using the trained model. 
Some researches~\cite{ref:device_cloud,ref:device_cloud2,zhang_device_cloud} have achieved great success on image classification tasks and recommender systems, demonstrating the potential value of the device-cloud collaboration scheme for DML.

Unfortunately, due to the heterogeneity of data distributions across devices, a primary cloud model trained on the data aggregated from all the devices might not generalize well and is typically less personalized to each particular device\cite{chen2021multi,li_multi,zhang2022towards,zhang2021unified,zhang2023dg,yuan2022domain,yuan2022label,lv2022personalizing}.
For instance, user data originating from devices of different geographical locales is potentially heterogeneous, leading to performance degradation for personalized DML~\cite{ref:device_cloud2,ref:specific_distribution_expert1,ref:specific_distribution_expert2}.
In other words, distribution shifts between the cloud and different devices require personalization of the cloud model before deployment on the device. 
To alleviate such \emph{data distribution shift} problem, Device Model Generalization (DMG) is needed to improve the generalization ability of a pre-trained model on a specific device. 
In this varying distribution setting, a commonly known DMG technique, as illustrated in Figure~\ref{fig1}(b), fine-tunes the pre-trained cloud model based on the current device data to mitigate the distribution shift issue, yielding an improvement of DMG for personalized learning\footnote{This paper studies the DMG problem in which primary models on the cloud and device have the same architecture but different parameters.}.

Despite their promising, fine-tuning based approaches may not be the true savior to resolve the DMG problems, due to the two key challenges summarized as follows.
(i) \textbf{Overfitting Issue and Annotation Demand.} The distribution of the real-time data input to the device is dynamic, and sometimes it may change drastically. 
For example, in the 
product recommendation task shown in Figure~\ref{fig1}, the user behavior sequence consists of dozens of recently clicked items by the users, which may exhibit different preferences in a short time snippet ($t$: \includegraphics[height=0.12in, width=0.12in]{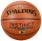} $\rightarrow$ $t+1$: \includegraphics[height=0.12in, width=0.12in]{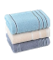} $\rightarrow$
$t+2$: \includegraphics[height=0.12in, width=0.12in]{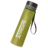}
). 
In order to achieve personalization, fine-tuning based DMG methods require re-training the model conditioned on those very limited samples with rapidly changing distributions on the device.  Such a learning paradigm may cause overfitting and performance degradation.
To make the matters worse, in vision tasks, the data on the device is generally not annotated.
To use the fine-tuning based methods, it is necessary to perform real-time annotation on the real-time data generated on the device. 
However, such frequent annotation is time-consuming and human-labor intensive to acquire for a device, which may not even be feasible in real-life applications.
(ii) \textbf{High Time-Resource Cost.} In addition to overfitting and extra annotation, on-device fine-tuning is time-consuming as it incurs numerous calculations on the gradients to update the model parameters, which is undesirable when the device applications typically have the real-time requirement constraint. It also consumes substantial amounts of device computing resources, thus leading to power consumption problem of smart devices. Therefore, these on-device training methods are not suitable for real-time DML in resource-constrained devices. Summing up, the premise of better DMG is to mitigate the aforementioned issues. The limitations require us to revisit the design of DMG solution for device-cloud collaboration.

In this paper, we propose a novel framework for DML, called  \textbf{\underline{D}}evice-clo\textbf{\underline{U}}d collaborative paramet\textbf{\underline{E}}rs genera\textbf{\underline{T}}ion  framework, (\textbf{DUET}) to address the aforementioned limitations. 
The core idea of our approach for DMG is to learn a device-specific model weight generator that dynamically adjusts from personalized data to solve a learning task. As shown in Figure~\ref{fig1}(c), our framework comprises the following parts: (1)  \textbf{Universal Meta Network} (UMN) first collects data from all the devices over a time period, labels them on the cloud server and then trains a primary model with these labeled data. For DMG, we divide the trained model into static layers and dynamic layers. The parameters of the static layers (backbone) are fixed, while the parameters of the dynamic layers (classifier) are dynamically generated based on device-specific real-time data in inference. (2) The \textbf{Personalized Parameters Generator} (PPG) leverages the \emph{HyperNetworks}~\cite{ref:hypernetworks} that efficiently share parameters across devices to generate separate classifier parameters for every device. Each device, with its unique real-time samples, passes as input to the designed PPG on the cloud to produce its personalized classifier weights. Then, the cloud will deliver the dynamic layer parameters to the device and enable real-time on-device inference with the personalized model. It is worth noting that the PPG is parallelly trained with UMN based on global data and only requires feedforward computation of local real-time data. The small time delay is due to the transmission of few data between the device and cloud, which produces a real-time DMG scheme. (3) We also propose a \textbf{Stable Weight Adapter} (SWA) which aims to generate stable parameters for dynamic layers by multiple PPGs due to the observation that one single PPG suffers from the performance oscillation problem. SWA measures the correlations between individualized trained PPGs, based on which a self-corrected adapter adaptively predict the optimal parameters of the primary model on the device. From the perspective of machine learning, the utilization of the correlation and similarity among related learning methods can be regarded as a form of inductive transfer. It can introduce the \emph{inductive bias}~\cite{baxter2000model} to make the combined learning method prefer the correct hypothesis, thereby improving the performance.  

In summary, as illustrated in Figure~\ref{fig1}(d), compared to the fine-tuning approaches that incur  \emph{high calculation cost and high annotation demand}, our proposed DUET entails \emph{zero calculation cost and zero annotation demand}. Our proposed DUET is arguably more pragmatic and suitable for real-time DMG. In this work, we make the following four key contributions:

\begin{itemize}
\item To the best of our knowledge, we are the first to incorporate the model parameter generation into device-cloud collaboration without expensive fine-tuning in on-Device Machine Learning.

\item We propose the personalized parameter generator which directly maps the device-specific data to model parameters for fast model personalization.

\item We design a stable weight adapter to reduce the performance oscillation of the dynamic model and further boost generalization across heterogeneous devices.

\item We conduct extensive experiments with various baselines on real-world benchmark datasets.
The results demonstrate the consistent superiority and generalizability of DUET. 
                                      
\end{itemize}

%% file: tex/2related_works.tex
\section{Related Work}
\label{sec:related_work}
\noindent\textbf{Lightweight Neural Network.}
The performance of traditional neural networks for different research tasks~\cite{chen2022ba,zhang2022boostmis,zhang2019frame,qin2020health} is already impressive. 
However, when the model is deployed on a device, the device's storage space and computing power have to be considered. 
Therefore, many lightweight CNN models~\cite{ref:squeezenet,ref:mobilenet,ref:mobilenetv2,ref:mobilenetv3, ref:shufflenet, ref:shufflenetv2,ref:efficientnet,ref:ghostnet} have been proposed in recent years. SqueezeNet~\cite{ref:squeezenet} reduces the number of parameters by extensively using fire modules with $1\times1$ convolutions. MobileNetV1~\cite{ref:mobilenet} decomposes traditional convolution kernels into depth-wise convolution kernels and point convolution kernels. MobileNetV2~\cite{ref:mobilenetv2} introduces inverted residuals and linear bottlenecks. MobileNetV3~\cite{ref:mobilenetv3} builds the network based on AutoML, manually fine-tunes the optimization to obtain the best network structure, and improves the performance and efficiency of the activation function. ShuffleNetV1~\cite{ref:shufflenet} uses channel shuffle to enhance information exchange between channel groups. ShuffleNetV2~\cite{ref:shufflenetv2} introduces channel split to improve inference speed. EfficientNet~\cite{ref:efficientnet} uses a neural network architecture (NAS) with a hybrid scaling method. GhostNet~\cite{ref:ghostnet} applies a linear transformation layer with fewer parameters to generate ghost feature maps. These models achieve good performance with small parameters and FLOPs, but the number of parameters still limits the performance and generalization ability.

\noindent\textbf{HyperNetwork.}
HyperNetwork~\cite{ref:hypernetwork_continual_learning,ref:hypernetwork_federated_learning,ref:hypernetwork_graph, ref:hypernetwork_initial,ref:hypernetwork_meta_learning,ref:hypernetworks,ref:hypernetwork_hyperstyle,ref:hypernetwork_hyperinverter} is a neural network that generates its parameters for another neural network. When HyperNetwork was first proposed by Ha et al.~\cite{ref:hypernetworks}, it achieved model compression by reducing the number of parameters the model needs to train. Subsequently, the research on HyperNetwork gradually increased. Oscar et al.~\cite{ref:hypernetwork_initial} studied parameter initialization for HyperNetwork. At the same time, the research of HyperNetwork is applied to various tasks, such as continual learning~\cite{ref:hypernetwork_continual_learning}, graph~\cite{{ref:hypernetwork_graph}}, meta-learning~\cite{ref:hypernetwork_meta_learning}, federated learning~\cite{ref:hypernetwork_federated_learning}, Etc. HyperNetwork-related research has mainly focused on generating different network parameters from different data inputs in the past two years. For example, HyperStyle~\cite{ref:hypernetwork_hyperstyle} and HyperInverter~\cite{ref:hypernetwork_hyperinverter} both use HyperNetwork to generate different decoder parameters for different images, thereby improving the quality of the reconstructed images. In our work, we adapt HyperNetwork to the device-cloud system with unique challenges arising from the problem setup.

%% file: tex/3method.tex
\section{Methodology}
\label{sec:method}
This section describes our proposed \textbf{D}evice-clo\textbf{U}d collabora\textbf{T}ive paramet\textbf{E}rs genera\textbf{T}ion  framework (\textbf{DUET}) for device model
generalization. We shall present each module and its training strategy.

\subsection{Problem Formulation}
In the problem of device model
generalization (DMG) in the device-cloud collaboration system, we have access to a set of devices $\mathcal{D}=\{d^{(i)}\}_{i=1}^{\mathcal{N}_d}$, each device with its personal i.i.d history samples  $\mathcal{S}_{H^{(i)}}=\{x^{(j)}_{H^{(i)}}, y^{(j)}_{H^{(i)}}\}_{j=1}^{\mathcal{N}_{H^{(i)}}}$ and real-time samples $\mathcal{S}_{R^{(i)}}=\{x^{(j)}_{R^{(i)}}\}_{j=1}^{\mathcal{N}_{R^{(i)}}}$ in current session, where $\mathcal{N}_{d}$, $\mathcal{N}_{H^{(i)}}$ and $\mathcal{N}_{R^{(i)}}$ represent the number of devices, history data and real-time data, respectively. The goal of DMG is to generalize a trained global cloud model $\mathcal{M}_g(\cdot;\Theta_g)$ learned from $\{\mathcal{S}_{H^{(i)}}\}_{i=1}^{\mathcal{N}_d}$ to each specific local device model $\mathcal{M}_{d^{(i)}}(\cdot;\Theta_{d^{(i)}})$ 
conditioned on real-time samples $\mathcal{S}_{R^{(i)}}$, where $\Theta_g$ and $\Theta_{d^{(i)}}$ respectively denote the learned parameters for the global and local models. 
\begin{equation}
\begin{aligned}
 \underbrace{\rm {\textbf{DUET}}}_{\rm{DMG\ Model}}: \underbrace{\mathcal{M}_{g}(\{\mathcal{S}_{H^{(i)}}\}_{i=1}^{\mathcal{N}_d};\Theta_g)}_{\rm{Global\ Cloud\ Model}}  \rightarrow
\underbrace{\mathcal{M}_{d^{(i)}}(\mathcal{S}_{R^{(i)}};\Theta_{d^{(i)}})}_{\rm{Local\ Device\ Model}}.
 \label{equ_1}
\end{aligned}
\end{equation}

\begin{figure*}[t]
  \centering
\includegraphics[width=0.9\linewidth]{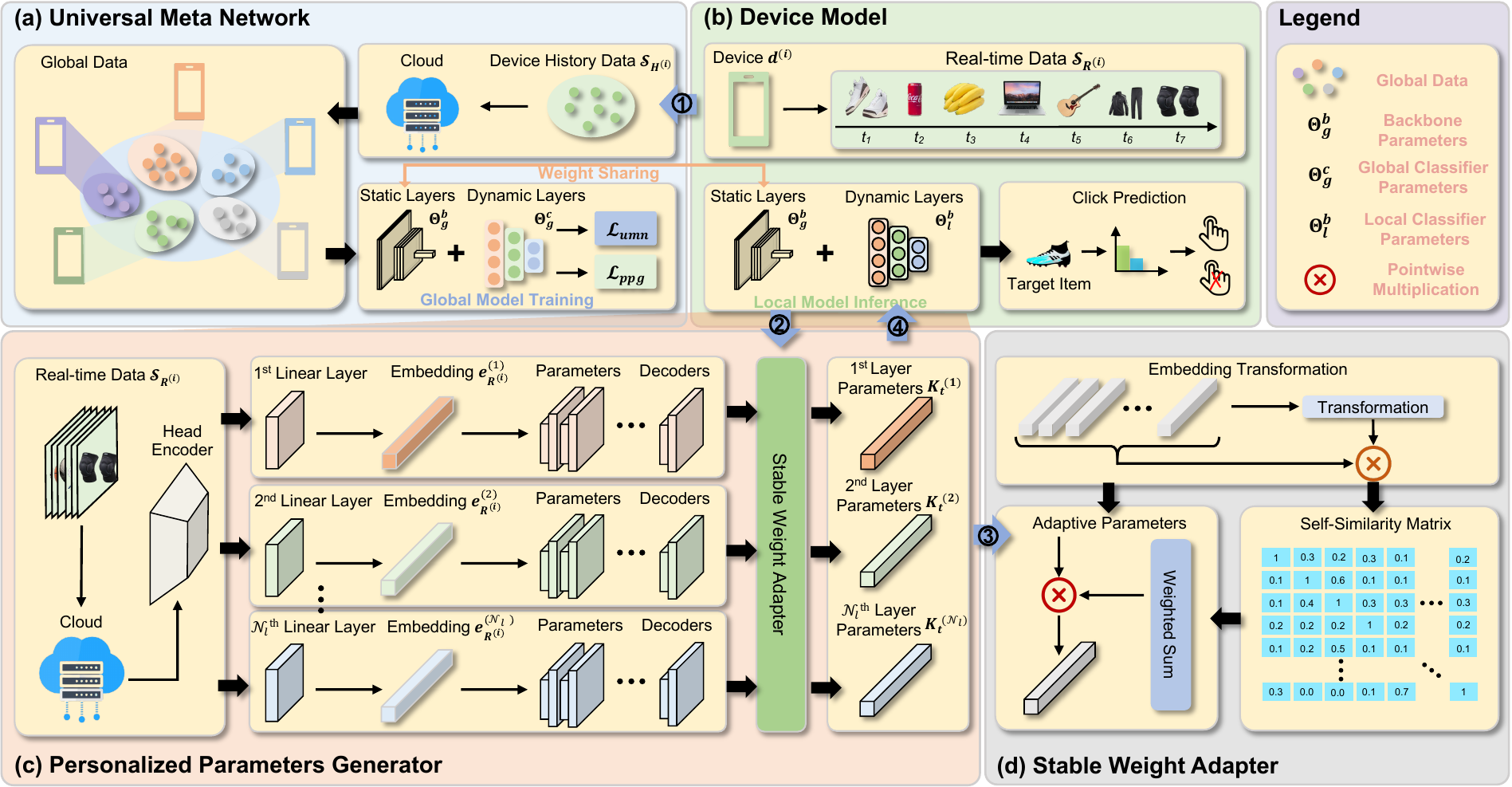}
\vspace{-0.3cm}
  \caption{Overview of the proposed \textbf{DUET}.
  The UMN is trained on the cloud that contains a backbone with parameters $\Theta_b$ and a classifier with parameters $\Theta_c$. The PPG is deployed on the cloud, which generates and delivers the personalized parameters ${\Theta}^c_l$ of dynamic layers for the device classifier based on the distribution of the real-time samples uploaded from the device. The SWA aims to reduce the performance oscillation of single PPG, accelerating the convergence and improving the prediction stability.}
  \label{fig:arch}
\vspace{-0.3cm}
\end{figure*}

Figure~\ref{fig:arch} illustrates the overview of our  DUET framework which consists of three modules to improve the generalization ability of the trained models on the device:
(a) \emph{Universal Meta Network} (UMN) aims to learn a global benchmark model based on the global data (in Sec.~\ref{subsec:umn});  (b) \emph{Personalized Parameters Generator} (PPG) that generates the network parameters for local device-specific model (in Sec.~\ref{subsec:ppg}); (c) \emph{Stable Weight Adapter} (SWA) presents the personalized parameter 
optimization strategy for robust DMG learning (in Sec.~\ref{subsec:swa}).

\subsection{Universal Meta Network.}
\label{subsec:umn}
In \textbf{Universal Meta Network} (UMN) (Figure~\ref{fig:arch}(a)), we train a primary model with a backbone and a classifier for the global cloud model development. Given a set of devices $\mathcal{D}=\{d^{(i)}\}_{i=1}^{\mathcal{N}_d}$ and their corresponding history data  $\mathcal{S}_{H^{(i)}}=\{x^{(j)}_{H^{(i)}}, y^{(j)}_{H^{(i)}}\}_{j=1}^{\mathcal{N}_{H^{(i)}}}$, the goal of the proposed UMN can thus be formulated as the following optimization problem:
 \begin{equation}
 \begin{aligned}
&\mathop{\rm min}_{\Theta_g^b, \Theta_g^c} \mathcal{L}_{umn}=\sum_{i=1}^{ \mathcal{N}_d} \sum_{j=1}^{ \mathcal{N}_{R^{(i)}}} 
D_{ce} (y^{(j)}_{H^{(i)}}, \Omega(x^{(j)}_{H^{(i)}};\Theta_g^b); \Theta_g^c)),
    \label{eq:umn}
\end{aligned}
\end{equation}
where $D_{ce}(·;\Theta_g^b)$ denotes the cross-entropy
between two probability distributions. $\Omega(x^{(j)}_{H^{(i)}};\Theta_g^b)$ is the backbone extracting features from sample $x^{(j)}_{H^{(i)}}$.
$\Theta_g^b$ and $\Theta_g^c$ are the learnable parameters for the backbone and classifier, respectively.

In our DMG setting,  we decouple the joint backbone and classifier training scheme as modeling the ``static layers'' and ``dynamic layers''  to achieve the personalized model generalization: 
\begin{itemize}
\item \textbf{Static Layers.} The backbone with $\Theta_g^b$ learned from global data can accurately map the user's behavior into the feature space. We fixed the backbone as ``static layers'' to generate a generalized representation for any given input concerning the global data distribution. 

\item \textbf{Dynamic Layers.} Depending on the user's behavior, the personalized samples obtained from a specific device are input to our proposed PPG to learn personalized classifier weights ${\Theta}_l^c$ . The improvement of personalized generalization can be achieved by just adjusting the classifier.
\end{itemize}

\subsection{Personalized Parameters Generator.}
The Personalized Parameters generator (PPG) can generate the dynamic parameters for the personalized classifier with ${\Theta}_l^c$ conditioned on the real-time samples from a specific device, which aims to improve the generalization to different distributions of data. We shall start by introducing the definition of \textbf{HyperNetwork}~\cite{ref:hypernetworks} and then propose our DMG mechanisms of parameters generation. 
\label{subsec:ppg}
\subsubsection{Retrospect of HyperNetwork}
\label{subsec:ppg_hp}
First, we will outline the procedure for using a \emph{HyperNetwork} to output the weights
of a feedforward convolutional network that performs the learning task. The \emph{HyperNetwork} regards the parameters $K^{(n)}$ for $n^{th}$ layer ($n \in \mathcal{N}_l$) of the CNN filter as a matrix of $\mathbb{R}^ {C_{in}f_w\times C_{out}f_h}$, where $\mathcal{N}_l$ is the depth of the main 
 network, the convolutional kernel contains $C_{in} \times C_{out}$ filters and each filter correspond the dimensions of $f_w \times f_h$. For $n^{th}$ layer, the \emph{HyperNetwork} is a two-layer MLP $g(\cdot;\Theta_p)$ with parameters $\Theta_p$ receives a layer embedding $z^{(n)} \in \mathbb{R}^{C_z}$ ($C_z \ll C_{in}f_w\times C_{out}f_h$) as input to and predicts $K^{(n)}$, which can be regard as the a matrix factorization scheme as follows:
 \begin{align}
    K^{(n)} = g(z^{(n)};\Theta_p), \forall n=  1, \cdots, \mathcal{N}_l.
\end{align}

In the \emph{training procedure}, $z^{(1)}\sim z^{(\mathcal{N}_l)}$ and $g(\cdot)$ are randomly initialized. As in a regular neural network, the network learns the mapping relationship between samples $x$ to $y$. 
Notably, the gradients are returned to $z^{(n)}$ and $g(\cdot)$ instead of $K^{(n)}$, which saves more space and computing power than storing parameters in $K^{(1)}\sim K^{(\mathcal{N}_l)}$. 
In the \emph{inference procedure}, $K^{(n)}$ is generated in blocks. The base convolution kernels $K^{'(n)}$ of all convolutional layers to be generated are set as convolution kernels with a dimension of ${C^{'}_{in}f^{'}_w\times C{'}_{out}f{'}_h}$. 
All convolutional layers to be generated need to conform that, (1) $f^{'}_w$ and $f{'}_h$ are required to be equal to $f_w$ and $f_h$, respectively. (2) ${C_{in}}$ and $C_{out}$ are integer multiples of ${C^{'}_{in}}$ and $C{'}_{out}$, respectively. 
Each base convolution kernel is generated by a base latent vector $z^{'(n)} \in \mathbb{R}^{C^{'}_z}$.
 The modular generation is below:
\begin{equation}
\begin{split}
    K^{(n)} &= 
    \left(
    \begin{array}{ccc}
      K^{'(n)}_{1,1}   &  \cdots  & K^{'(n)}_{1,j}\\
      \vdots     & \ddots  &  \vdots  \\
      K^{'(n)}_{i,1} &  \cdots  &  K^{'(n)}_{i,j}
    \end{array}
    \right)  
    =g
    \left(
    \begin{array}{ccc}
      z^{'(n)}_{1,1}   &  \cdots  & z^{'(n)}_{1,j}\\
      \vdots     & \ddots  &  \vdots  \\
      z^{'(n)}_{i,1} &  \cdots  &  z^{'(n)}_{i,j}
    \end{array}
    \right) \\ 
    & \triangleq g(z^{(n)}),  i=\frac{C_{in}}{C^{'}_{in}}, j=\frac{C_{out}}{C^{'}_{out}}, \forall  n=  1, \cdots, \mathcal{N}_l.
\end{split}
\label{eq:hypernetwork_modular_generation}
\end{equation}

After retrospecting  characteristics of \emph{HyperNetwork}, 
which seems naturally suitable for learning a diverse set of personalized models. As \emph{HyperNetwork} dynamically generates target networks conditioned on the input embeddings, \emph{i.e.}, the ``Dynamic Layers'' of the primary model can be modeled by \emph{HyperNetwork}. 
However, our observation (in Appendix) indicates that directly utilizing the \emph{HyperNetwork} may not satisfactorily resolve the DMG problem for two key reasons:
\begin{itemize}
\item \textbf{Weak Correlation.} The original \emph{HyperNetwork} uses a random latent vector $z$ to initialize the model that lacks the strong correlation between parameters generation and a specific device, which may yield a performance decay.

\item \textbf{Unstable Prediction.} The empirical experiments indicate that the performance of \emph{HyperNetwork} is intuitively unstable during training and inference, mainly because a single \emph{HyperNetwork} is hard to measure the parameters.
\end{itemize}

To this end,
we carefully design the Personalized Parameters Generator (PPG) and Stable Weight Adapter (SWA) to deal with the above limitations. 

\subsubsection{Device-specific Parameters Generation.}
Considering the \emph{Weak Correlation} between \emph{HyperNetwork} and the random latent vector $z$, as shown in Figure~\ref{fig:arch}(b), we propose to model the ``Dynamic Layers'' of the primary model by replacing the $z$ with specific samples from devices in inference. 
Further, to satisfy the architecture consistency of pre-trained classifier and ``Dynamic Layers'', we develop a hierarchical \emph{HyperNetworks} to generate its parameters.

\begin{sloppypar} 
\emph{In device inference}, we use the real-time samples $\mathcal{S}_{R^{(i)}}=\{x^{(j)}_{R^{(i)}}\}_{j=1}^{\mathcal{N}_{R^{(i)}}}$ in each session to generate the model parameters. 

To generate the parameters for $n^{th}$ layer of ``Dynamic Layers'' in the primary model, we develop a layer encoder to represent the $n^{th}$ layer parameters as an embedding $\boldsymbol{e}^{(n)}_{R^{(i)}}$. To model relationships of different layers, instead of constructing the one-to-one encoder-layer correspondence, the $\boldsymbol{e}^{(n)}_{R^{(i)}}$ share one encoder neck but use different linear layers to change the real-time data features.
\end{sloppypar}
\begin{align}
\label{eq:lightweight_encoder}
    \boldsymbol{e}^{(n)}_{R^{(i)}} = {L}^{(n)}_{\rm{layer}}({E}_{\rm{share}} (\mathcal{S}_{R^{(i)}})), \forall n=  1, \cdots, \mathcal{N}_l,
\end{align}
where ${E}_{\rm{share}}(\cdot)$ represents the shared encoder neck. ${L}^{(n)}_{\rm{layer}}(\cdot)$ is a linear layer used to adjust the output of ${E}_{\rm{share}}(\cdot)$ to the $n^{th}$ dynamic layer features.

We treat it as a matrix $K^{(n)} \in \mathbb{R}^{N_{in}\times N_{out}}$, where $N_{in}$ and $N_{out}$ represent the number of input neurons and output neurons of the $n^{th}$ FCL, respectively. 
Then we use the generator $g(\cdot)$ to convert the real-time data features into parameters of the primary models by $K^{(n)}_{R^{(i)}} = g^{(n)}(\boldsymbol{e}^{(n)}_{R^{(i)}})$.
Specifically, we input $\boldsymbol{e}^{(n)}_{R^{(i)}}$ into the following two MLP layers to generate parameters according to the consistent structure of ``Dynamic Layers'' of the primary model. 
\begin{align}
\label{eq:kernal_generation_detail}
\begin{gathered}
    \mathcal{\boldsymbol{w}}^{(n)}_{R^{(i)}} = (W_1\boldsymbol{e}^{(n)}_{R^{(i)}} + B_1)W_2 + B_2, \\
    K^{(n)}_{R^{(i)}} = \mathcal{\boldsymbol{w}}^{(n)}_{R^{(i)}} + \mathcal{\boldsymbol{b}}^{(n)}_{R^{(i)}},
\end{gathered}
\end{align}
where weights of the two MLP layers are denoted by $W_1$ and $W_2$, respectively. $B_1$ and $B_2$ indicate the biases.

\emph{In cloud training}, all layers of the PPG are optimized together with the static layers of the primary model that are conditioned on the global history data
$\mathcal{S}_{H^{(i)}}=\{x^{(j)}_{H^{(i)}}, y^{(j)}_{H^{(i)}}\}_{j=1}^{\mathcal{N}_{H^{(i)}}}$,  instead of optimizing the static layers of the primary model first and then optimizing the PPG. 
The PPG loss function $\mathcal{L}_{ppg}$ is defined as follows:
\begin{align}
    \label{eq:loss_func}
\!\!\!\! &\mathop{\rm min}_{\Theta_g^b, \Theta_p}\!\! \mathcal{L}_{ppg}=\!\!\!\!\sum_{i=1}^{ \mathcal{N}_d} \sum_{j=1}^{ \mathcal{N}_{R^{(i)}}} \!\!
\gamma^t D_{ce} (y^{(j)}_{H^{(i)}}, \Omega(x^{(j)}_{H^{(i)}};\Theta_g^b); g(\boldsymbol{e}^{(n)}_{R^{(i)}};\Theta_p))),   \!\!
\end{align}
where $\gamma$ is a hyperparameter used to adjust the training. When the $\gamma$ is closer to 1, the PPG considers all samples in each session as equally important. Otherwise, PPG pays more attention to the earlier samples in each session. 

Here we use a special group-wise convolution in PPG, so that the entire framework can generate parameters for the primary models during training and inference in parallel. This greatly improves PPG training and inference efficiency and makes it easier to deploy in real environments.

\subsubsection{Stable Weight Adapter.}
\label{subsec:swa}

As described in Sec.~\ref{subsec:ppg_hp}, directly using \emph{HyperNetwork} often produces large oscillations during learning, and thus yields \emph{Unstable Prediction}. 
To resolve this issue,
we propose the Stable Weight Adapter (SWA) module to improve the prediction stability.

\begin{table*}[t]
  \caption{
  Performance comparison of the proposed method and baselines on sequential recommendation datasets (\texttt{Movielens-1M} and \texttt{Movielens-100k}). \textbf{\emph{$LR^*$}} indicates that the DMG models  require the label information of real-time data from device.  $\uparrow$ and $\downarrow$ respectively indicate a larger and smaller score has better performance.  Acronym notations of baselines can be found in Sec.~\ref{subsec:experiment_baseline}.
  We color each row as the \colorbox{myred}{\textbf{best}}, \colorbox{myorange}{\textbf{second best}}, and \colorbox{myyellow}{\textbf{third best}}.} 
  \label{tab:experiment_rs}
  \centering
 \renewcommand{\arraystretch}{1.2}
 \resizebox{0.98\textwidth}{!}{
{
    \begin{tabular}{c|c|c|c|c|c|c|c|c|c|c|c|cc}
    \toprule[2pt]
     \multirow{2}{*}{\textbf{Baselines}}&\multirow{2}{*}{\textbf{DMG Methods}} & \multirow{2}{*}{\textbf{\emph{$LR^*$}}} & \multicolumn{4}{c|}{\texttt{Movielens-1M Dataset}} & \multirow{2}{*}{\textbf{Time Delay}} & \multicolumn{4}{c|}{\texttt{Movielens-100k Dataset}} & \multirow{2}{*}{\textbf{Time Delay}} \\
     \cline{4-7}\cline{9-12}
    &
    & &\textbf{AUC(mean)} $\uparrow$ & \textbf{AUC(std)} $\downarrow$ & \textbf{Param.} & \textbf{FLOPs} & & \textbf{AUC(mean)} $\uparrow$ & \textbf{AUC(std)} $\downarrow$ & \textbf{Param.} & \textbf{FLOPs} & \\
    \midrule[1pt]
    \midrule[1pt]
    \multirow{4}{*}{DIN~\cite{ref:din}} & - && {0.9077} & \colorbox{myred}{{0.0006}} & \multirow{4}{*}{896.83K} & \multirow{4}{*}{1.82M} & 0 & {0.8348} & \colorbox{myorange}{0.0045} & \multirow{4}{*}{271.83K} & \multirow{4}{*}{0.60M}  & 0\\ 
    \cline{2-5} \cline{9-10} \cline{8-8} \cline{13-13}
    
     
    \multirow{4}{*} & Fine-tuning &\Checkmark& \colorbox{myyellow}{{0.9080}} & \colorbox{myred}{0.0006} & & & $\geq$60,000ms & \colorbox{myyellow}{0.8429} & \colorbox{myorange}{0.0045} &  &  & $\geq$60,000ms\\

    \cline{2-5} \cline{9-10} \cline{8-8} \cline{13-13}
    
    \multirow{4}{*} & DUET (w/o SWA) && \colorbox{myred}{{0.9233}} & \colorbox{myyellow}{0.0008} & & & \multirow{2}{*}{$\geq$0.34ms} & \colorbox{myorange}{0.8560} & \colorbox{myred}{{0.0030}} & & & \multirow{2}{*}{$\geq$0.15ms}\\
    \cline{2-5} \cline{9-10}
    
    \multirow{4}{*} & \cellcolor{gray!40}\textbf{DUET}  &\cellcolor{gray!40}& \cellcolor{gray!40}\colorbox{myorange}{{0.9230}} & \cellcolor{gray!40}\colorbox{myorange}{{0.0007}} & & & & \cellcolor{gray!40}\colorbox{myred}{{{0.8581}}} & \cellcolor{gray!40}\colorbox{myyellow}{{0.0055}} & & & \\ 
    \midrule[1pt]

    \multirow{4}{*}{SASRec~\cite{ref:sasrec}} & - && \colorbox{myyellow}{0.9280} & \colorbox{myyellow}{0.0007} & \multirow{4}{*}{888.63K} & \multirow{4}{*}{1.99M} & 0 & \colorbox{myorange}{0.8721} & \colorbox{myorange}{0.0026} & \multirow{4}{*}{263.63K} & \multirow{4}{*}{0.77M} & 0\\ \cline{2-5} \cline{9-10} \cline{8-8} \cline{13-13}
    
     
    \multirow{4}{*} & Fine-tuning &\Checkmark& {0.9279} & \colorbox{myorange}{0.0006} &  &  & $\geq$60,000ms & \colorbox{myyellow}{0.8719} & \colorbox{myyellow}{0.0027} &  & & $\geq$60,000ms\\ \cline{2-5} \cline{9-10} \cline{8-8} \cline{13-13}
    
    \multirow{4}{*} & DUET (w/o SWA) && \colorbox{myorange}{0.9313} & \colorbox{myyellow}{0.0007} & & & \multirow{2}{*}{$\geq$0.34ms} & \colorbox{myorange}{0.8721} & \colorbox{myyellow}{0.0027} & & & \multirow{2}{*}{$\geq$0.15ms}\\
    \cline{2-5} \cline{9-10}
    
    \multirow{4}{*} & \cellcolor{gray!40}\textbf{DUET} &\cellcolor{gray!40}& \cellcolor{gray!40}\colorbox{myred}{{{0.9326}}} & \cellcolor{gray!40}\colorbox{myred}{{{0.0003}}} & & & & \cellcolor{gray!40}\colorbox{myred}{{{0.8723}}} & \cellcolor{gray!40}\colorbox{myred}{{{0.0009}}} & & & \\ 
    
    \midrule[1pt]
    \multirow{4}{*}{GRU4Rec~\cite{ref:gru4rec} }& - && \colorbox{myyellow}{0.9279} & \colorbox{myyellow}{0.0016} & \multirow{4}{*}{886.62K} & \multirow{4}{*}{1.94M}  & 0 & \colorbox{myyellow}{0.8723} & \colorbox{myred}{{0.0017}} & \multirow{4}{*}{261.62K} & \multirow{4}{*}{0.72M}  & 0\\ \cline{2-5} \cline{9-10} \cline{8-8} \cline{13-13}

    
    \multirow{4}{*} & Fine-tuning &\Checkmark & \colorbox{myorange}{0.9286} & \colorbox{myorange}{0.0014} &  & & $\geq$60,000ms & {0.8711} & \colorbox{myorange}{0.0019} &  &  & $\geq$60,000ms \\ \cline{2-5} \cline{9-10} \cline{8-8} \cline{13-13}
    
    \multirow{4}{*} & DUET (w/o SWA) && \colorbox{myred}{{0.9311}} & \colorbox{myorange}{0.0014} & & & \multirow{2}{*}{$\geq$0.34ms} & \colorbox{myorange}{0.8751} & \colorbox{myyellow}{0.0055} & & & \multirow{2}{*}{$\geq$0.15ms}\\ \cline{2-5} \cline{9-10}
    
    \multirow{4}{*} & \cellcolor{gray!40}\textbf{DUET} &\cellcolor{gray!40}& \cellcolor{gray!40}\colorbox{myred}{{{0.9311}}} & \cellcolor{gray!40}\colorbox{myred}{{{0.0004}}} & & & & \cellcolor{gray!40}\colorbox{myred}{{{0.8755}}} & \cellcolor{gray!40}\colorbox{myred}{{{0.0017}}} & & & \\
    \bottomrule[2pt]
    \end{tabular}
    }
}
\end{table*}

\begin{table}[t]
  \caption{
  Performance comparison of the proposed method and baselines on facial expression recognition datasets (\texttt{CK+}). } 
  \label{tab:experiment_cv}
  \centering
 \renewcommand{\arraystretch}{1.2}
  \resizebox{0.48\textwidth}{!}{

    \begin{tabular}{c|c|c|c|c|c|c}
    \toprule[2pt]
     \multirow{2}{*}{\textbf{Baselines}}&\multirow{2}{*}{\textbf{DMG Methods}} &\multicolumn{4}{c|}{\texttt{CK+ Dataset}} & \multirow{2}{*}{\textbf{Time Delay}} \\
     \cline{3-6}
    & & \textbf{AUC(mean)} $\uparrow$ & \textbf{AUC(std)} $\downarrow$ & \textbf{Param.} & \textbf{FLOPs} & \\
    \midrule[1pt]
    \midrule[1pt]
    \multirow{3}{*}{\makecell{MobileNetV3\\-Large~\cite{ref:mobilenetv3}}
    } & - &   \colorbox{myyellow}{76.97} & \colorbox{myorange}{2.76} & \multirow{3}{*}{2.69M} & \multirow{3}{*}{0.27B}  & 0\\ 
    \cline{2-4} \cline{7-7}

    \multirow{3}{*} & DUET (w/o SWA)   & \colorbox{myorange}{79.49}  & \colorbox{myyellow}{2.84} &  &   & \multirow{2}{*}{$\geq$51ms} \\ \cline{2-4}

    \multirow{3}{*} & \cellcolor{gray!40}\textbf{DUET}   & \cellcolor{gray!40}\colorbox{myred}{{82.45}} & \cellcolor{gray!40}\colorbox{myred}{{2.67}} & &  & \\ 
    \midrule[1pt]

 
     \multirow{3}{*}{\makecell{MobileNetV3\\-Small~\cite{ref:mobilenetv3}}} & -  & \colorbox{myyellow}{67.88} & \colorbox{myorange}{2.15} & \multirow{3}{*}{1.24M} & \multirow{3}{*}{0.06B}  & 0\\ \cline{2-4} \cline{7-7}

    \multirow{3}{*} & DUET (w/o SWA)  & \colorbox{myorange}{71.82} & \colorbox{myyellow}{2.91} & &  &  \multirow{2}{*}{$\geq$31ms} \\
    \cline{2-4}
    \multirow{3}{*} & \cellcolor{gray!40}\textbf{DUET} & \cellcolor{gray!40}\colorbox{myred}{{{75.15}}} & \cellcolor{gray!40}\colorbox{myred}{{{1.41}}} & &  & \\ 
    
    \bottomrule[2pt]
    \end{tabular}
    
}
\end{table}
First, we develop $\mathcal{N}_p$ PPGs to generate a set of parameters rather than a single generator are trained in the same way, denoted by $W^{'}_1$. Then splicing multiple $W_1^{'}$ into a matrix, denoted as $W^{'}_1=\{W^{'}_{1,1},W^{'}_{1,2}, ... , W^{'}_{1,m}\}$. $W^{'}_{i,j}$ denoted the $i$-th MLP layer of the $j$-th generator. Then we can get the similarity between $W^{'}_{1,i}$ and $W^{'}_{1,j}$, thus a self-similarity matrix $S$ of dimension $m\times m$ is obtained by $S = W^{'}_{1} * (W^{'}_{1})^T$.

\begin{sloppypar}
Summing $S$ by row, we can get the weight vector $\boldsymbol{p}^{'}=\{p^{'}_1, p^{'}_2, ..., p^{'}_m\}$ with dimension $m\times 1$. Among them, $p_i$ can be regarded as the importance of $W^{'}_{1,i}$ in the multiple generators. We also set temperature to adjust the final weight vector $\boldsymbol{p}$,
\end{sloppypar}
\vspace{-0.3cm}
\begin{align}
p_i = {\rm{Softmax}}(\frac{p^{'}_i/\tau}{\sum^{m}_{j=1} p^{'}_j})
\end{align}
\vspace{-0.3cm}

Then we can calculate the final $W_1$ and $W_2$ like,
\begin{align}
\label{eq:encoder_ensemble}
W_1 = \sum^m_{i=1}{p_i * W^{'}_{1,i}}
\end{align}

Finally, we use Eq.~(\ref{eq:encoder_ensemble}) to get $W_1$ and $W_2$, and get the model parameters after replacing $W_1$ and $W_2$ in Eq.~(\ref{eq:kernal_generation_detail}).

%% file: tex/4experiment.tex
\section{Experiments}
\label{sec:experiments}
We conduct a range of sequential recommendation and facial expression recognition experiments on three public datasets to demonstrate the effectiveness of the proposed DUET framework.
\subsection{Experimental Setup}
\subsubsection{Datasets.}
\begin{sloppypar} \noindent \textbf{Sequential Recommendation.}
We evaluate DUET on \texttt{Movielens-1M}
and \texttt{Movielens-100k} \footnote{\url{http://grouplens.org/datasets/movielens/}\label{fn:movielens}}, 
two widely used public benchmarks in the recommendation tasks. Following conventional practice, all user-item pairs in the dataset are treated as positive samples. In the training and test sets, the user-item pairs that do not exist in the dataset are sampled at 1:4 and 1:100, respectively, as negative samples~\cite{ref:din,ref:sasrec,ref:gru4rec}. 
\textbf{Facial Expression Recognition.} We evaluate our method on \texttt{CK+} \footnote{\url{https://www.jeffcohn.net/Resources/}}
~\cite{ref:ck,ref:ck+}. \texttt{CK+} is a facial expression recognition dataset containing 593 videos of 123 people, of which 327 videos are annotated with expressions. The labels contain seven basic emotions: anger, contempt, disgust, fear, happiness, sadness, and surprise. To simulate a real device-cloud collaborative environment, we treat each video as a session and keep each video's first and last three frames. Since each video records a person from no expression to an exaggerated expression consistent with the label, we set the label of the first three frames as natural and the label of the last three frames as the emotion label of the video. 
\end{sloppypar}

\subsubsection{Baselines.} 
\label{subsec:experiment_baseline}

\begin{sloppypar}
In the sequential recommendation task, DIN~\cite{ref:din}, SASRec~\cite{ref:sasrec}, and GRU4Rec~\cite{ref:gru4rec}, three of the most widely used methods in the academia and industry, are chosen as the baselines. In the facial expression recognition task, we choose MobileNetV3~\cite{ref:mobilenetv3} as the baseline, which is one of the most popular lightweight networks.
\end{sloppypar}

\subsubsection{Implementation Details.} 
\label{subsec:experiment_implementation_details}
\textbf{Training Procedure}. 
When training DUET, in the sequential recommendation task, we input the most recent click sequence as the real-time samples to PPG to get the dynamic layers' parameters. Then we update the obtained parameters to the primary model's dynamic layers and model the mapping relationship between samples and labels in this session. 
In the Facial Expression Recognition task, the first frame of each video is set to a real-time sample. After updating the parameters similar to the above process, we model the mapping relationship between samples and labels in this video. Gradients are passed back to UMN's static layers and PPG.

\noindent
\textbf{Inference Procedure}. 
The inference process of baselines is performed on the device.
Fine-tuning based DMG methods need to fine-tune the base model with real-time data and then make inferences on the device. 
In the inference process of DUET, the device first uploads real-time samples to the cloud. 
PPG then generates the parameters of dynamic layers in UMN according to the distribution information of real-time data samples and sends them to the device. 
The updated parameters of dynamic layers will be sent to devices, and make inferences together with static layers.
The device only requests model parameters at the beginning of each session/video.
In the actual deployment, the recommendation task regards opening the APP and refreshing the page as the beginning of a session. The vision task's content at a fixed time interval is regarded as a video. Table~5 in the Appendix shows the hyperparameters and training schedules of DUET on the three datasets.

\begin{figure}[t]
  \centering
\includegraphics[width=0.93\linewidth]{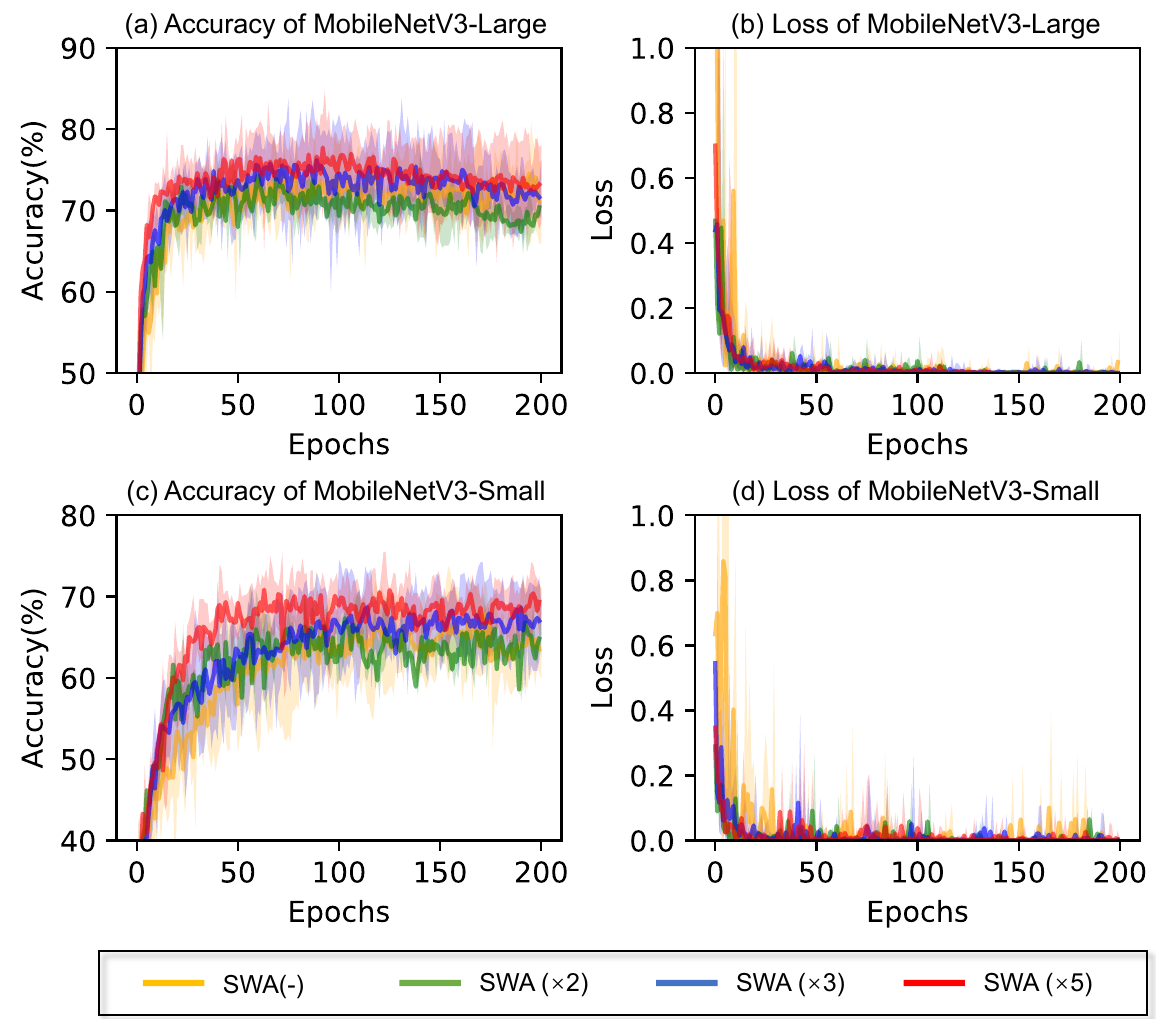}
\vspace{-0.12cm}
  \caption{Effects of the number of SWAs in training.}
  \label{fig:training_loss_curve_moe_num}
\end{figure}

\begin{figure}[t]
  \centering
\includegraphics[width=0.75\linewidth]{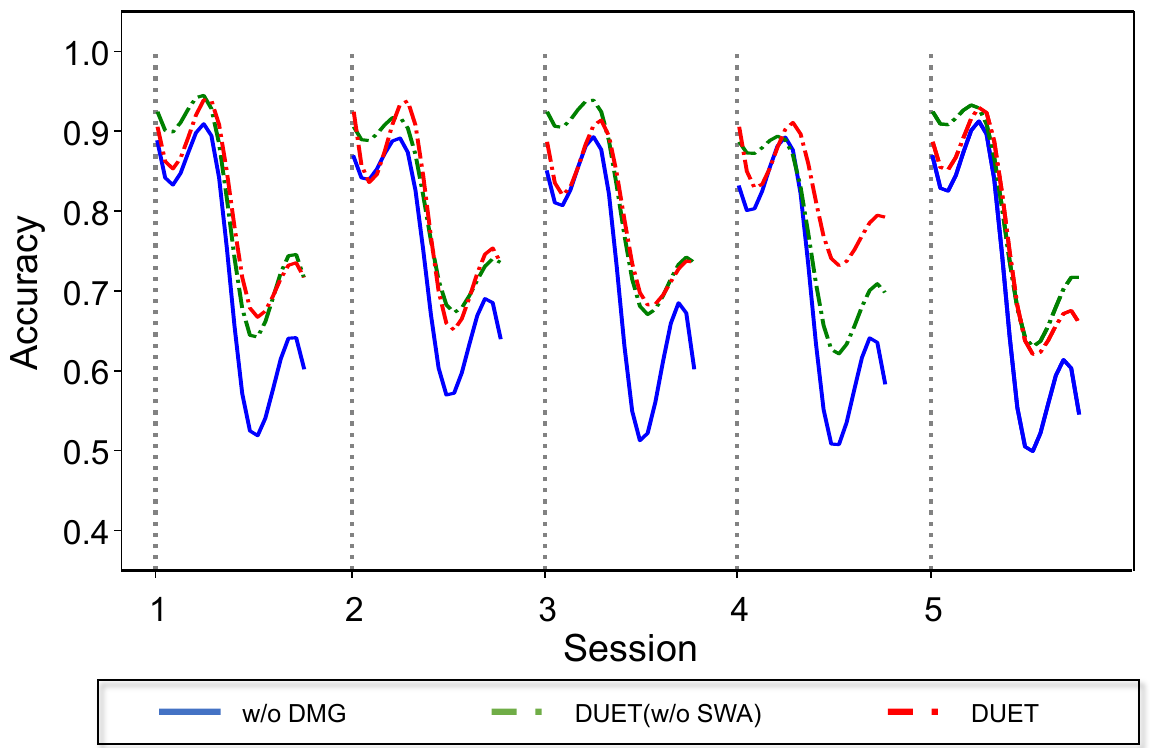}
\vspace{-0.12cm}
  \caption{Performance comparison on each session.}
  \label{fig:before_and_after_updating}
\end{figure}

\begin{figure}
  \centering
\includegraphics[width=0.93\linewidth]{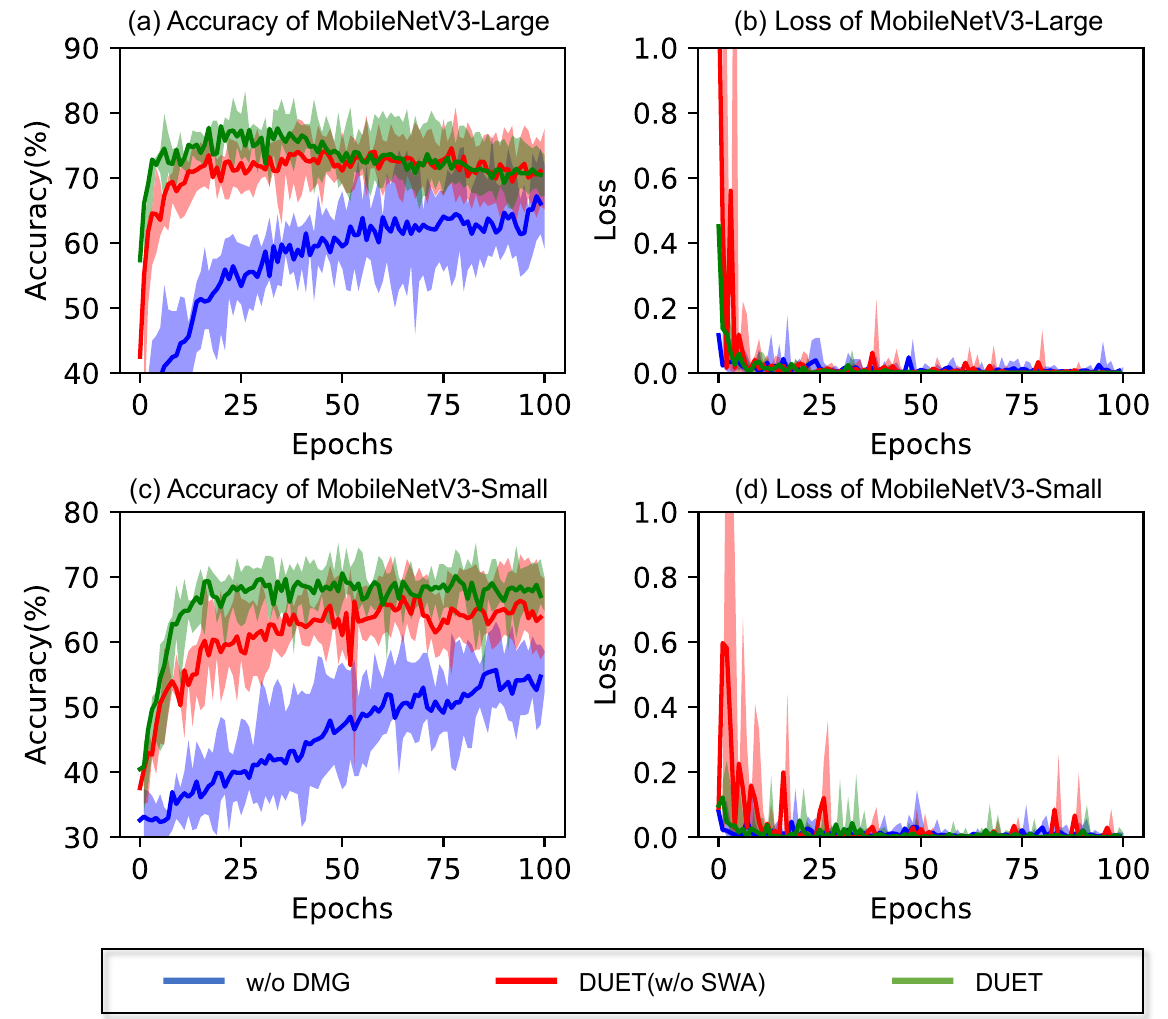}
\vspace{-0.12cm}
  \caption{Training visualization of DUET and baselines.}
  \label{fig:acc_loss_error_bar}
\end{figure}

\begin{figure}[t]
  \centering
\includegraphics[width=0.9\linewidth]{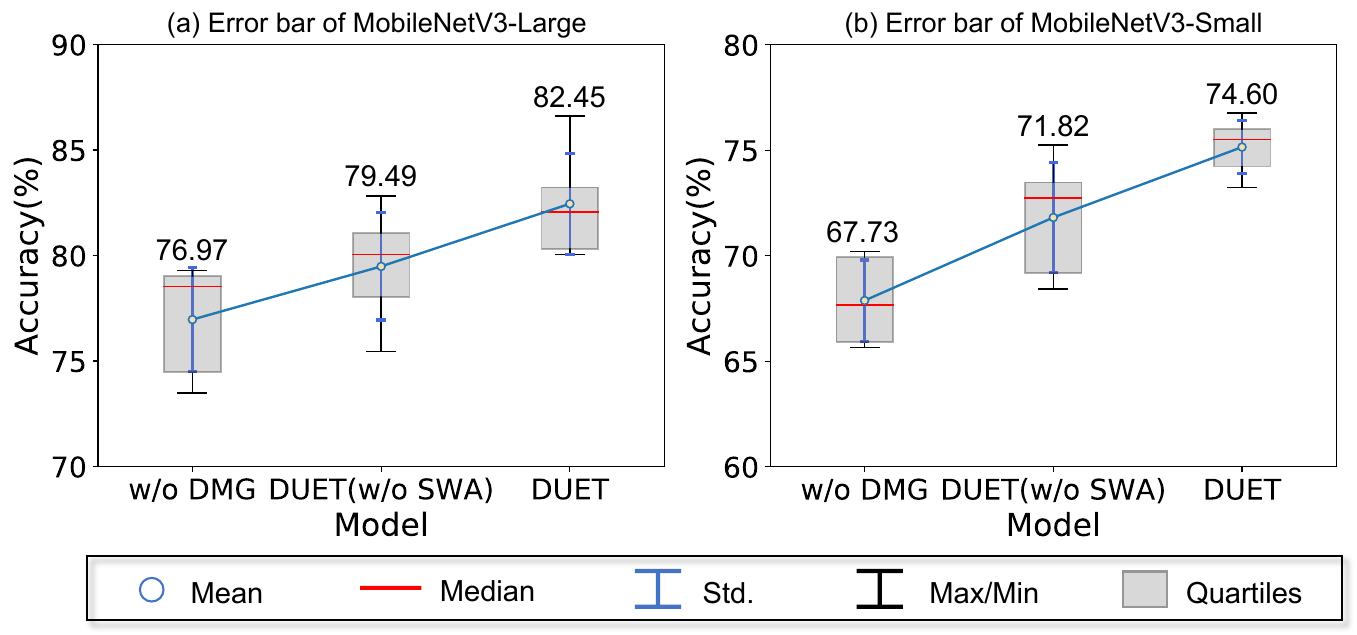}
\vspace{-0.12cm}
  \caption{Detailed performance comparison of DUET and baselines.}
  \label{fig:performance_boxplot}
\end{figure}
\vspace{-0.13cm}
\subsection{Experimental Results.}

\noindent \textbf{Results of Sequential Recommendation.} 
Table~\ref{tab:experiment_rs} summarizes the quantitative results of our framework and other DMG methods on \texttt{Movielens-1M} and \texttt{Movielens-100k} dataset. From this table, we have the following findings: 
(1) Almost all DMG models can improve the baseline's performance (AUC (std)) across the two datasets, which demonstrates the application value of model generalization on the device.
(2) The effect of model fine-tuning is insignificant, and we observe performance degradation in some cases, \emph{e.g.}, fine-tuning for baseline SASRec (Row. 6). This phenomenon is reasonable as the fine-tuning model may encounter the over-fitting issue when trained on limited real-time data. In addition, it also causes a high-time delay, which is impractical for applications on the device. 
(3) DUET and its variant DUET (w/o SWA) both outperform fine-tuning-based DMG model by a large margin, \emph{e.g.}, \textbf{\underline{DUET (w/o SWA) improves 0.0153 AUC (mean) ( Row. 3) and}}
\textbf{\underline{DUET improves 0.0150 AUC (mean) (Row. 4)}} using DIN baseline, respectively. Notably, it enables a real-time recommendation with an extremely low time delay. 
(4) DUET (w/o SWA) and DUET produce a similar performance in all datasets. However, Table~\ref{tab:experiment_rs} suggests that DUET consistently maintains a stable performance across all  the baselines, \emph{i.e.}, the AUC (std) of DUET  is smaller than DUET (w/o SWA), \emph{e.g.},  Row. 7 \emph{vs} Row. 8. These results indicate that the SWA can bring improved robustness for the DMG task. 

\begin{table*}[t]
  \caption{Time delay caused by device-cloud communication, $\Uparrow$ and $\Downarrow$ represent upload (device $\rightarrow$ cloud) and download (cloud $\rightarrow$ device), respectively. Data transmission \textbf{||} embedding transmission indicates the different upload settings.
 }
\vspace{-0.087cm}
  \label{tab:delivery_delay}
  \centering
\resizebox{0.97\textwidth}{!}{
    \begin{tabular}{c|c|c|c|c|c|c}
    \toprule[2pt]
    \textbf{Datasets} & \textbf{Models} & \textbf{Size} & \textbf{4G: 5MB/s}  & \textbf{4G: 15MB/s} & \textbf{5G :50MB/s} & \textbf{5G: 100MB/s} \\
    \midrule[1pt]
    \midrule[1pt]
    \multirow{3}{*}{\makecell[c]{\texttt{CK+}}} & \makecell{MobileNetV3-\\Large} & \makecell{$\Uparrow$: 0.14MB || 0.25KB \\ {$\Downarrow$: 5.31MB} } & \makecell{$\Uparrow$: 0.03s || 0.05ms\\$\Downarrow$: 1.06s}
    & \makecell{$\Uparrow$: 0.01s || 0.017ms\\ $\Downarrow$: 0.35s} & 
    \makecell{$\Uparrow$: 0.003s || 0.005ms\\ $\Downarrow$: 0.10s} & 
    \makecell{$\Uparrow$: 0.001s || 0.003ms\\ $\Downarrow$: 0.05s} \\ \cline{2-7} 
    \multirow{3}{*} & \makecell{MobileNetV3-\\Small} & 
    \makecell{$\Uparrow$: 0.14MB || 0.25KB\\ $\Downarrow$: 3.06MB} &	
    \makecell{$\Uparrow$: 0.03s || 0.05ms\\ $\Downarrow$: 0.61s} &	
    \makecell{$\Uparrow$: 0.01s || 0.017ms\\ $\Downarrow$: 0.20s} &	
    \makecell{$\Uparrow$: 0.003s || 0.005ms\\ $\Downarrow$: 0.06s} & 
    \makecell{$\Uparrow$: 0.001s || 0.003ms\\ $\Downarrow$: 0.03s} \\
    \midrule[1pt]
    \multirow{3}{*}{\makecell[c]{\texttt{Movielens-1M}}} & DIN & 
    \multirow{3}{*}{\makecell{$\Uparrow$: 25.63KB || 0.25KB\\ $\Downarrow$: 8.06KB}} & 
    \multirow{3}{*}{\makecell{$\Uparrow$: 5.13ms || 0.05ms\\ {$\Downarrow$: 1.60ms}}} & 
    \multirow{3}{*}{\makecell{$\Uparrow$: 1.71ms || 0.017ms\\ {$\Downarrow$: 0.53ms}}} & 
    \multirow{3}{*}{\makecell{$\Uparrow$: 0.51ms || 0.005ms\\ {$\Downarrow$: 0.16ms}}} & 
    \multirow{3}{*}{\makecell{$\Uparrow$: 0.26ms || 0.003ms\\ {$\Downarrow$: 0.08ms}}} \\ 
    \cline{2-2}
     & SASRec &  &  &  &  &  \\
    \cline{2-2}
     & GRU4Rec &  &  &  &  &  \\
    \midrule[1pt]
     \multirow{3}{*}{\makecell[c]{\texttt{Movielens-100k}}} & DIN & \multirow{3}{*}
     {\makecell{$\Uparrow$: 7.32KB || 0.25KB\\ {$\Downarrow$: 8.06KB}}} & 
     \multirow{3}{*}{\makecell{$\Uparrow$: 1.46ms || 0.05ms\\ {$\Downarrow$: 1.60ms}}} & 
     \multirow{3}{*}{\makecell{$\Uparrow$: 0.49ms || 0.017ms\\ {$\Downarrow$: 0.53ms}}} & 
     \multirow{3}{*}{\makecell{$\Uparrow$: 0.15ms || 0.005ms\\ {$\Downarrow$: 0.16ms}}} & 
     \multirow{3}{*}{\makecell{$\Uparrow$: 0.07ms || 0.003ms\\ {$\Downarrow$: 0.08ms}}} \\ 
    \cline{2-2}
     & SASRec &  &  &  &  &  \\
    \cline{2-2}
     & GRU4Rec &  &  &  &  &  \\
    \bottomrule[2pt]
    \end{tabular}
    }
\end{table*}
\begin{table}[t]
    \caption{Effect of the number of SWAs on performance. The best results are highlighted in bold.}
\vspace{-0.088cm}
    \centering
\resizebox{0.45\textwidth}{!}{
    \begin{tabular}{c|c|c|c|c}
    \toprule[2pt]
    {\textbf{Baselines}}&{\textbf{DMG Methods}} & \textbf{$\mathcal{N}_p$} & \textbf{Accuracy(\%)} & \textbf{Std(\%)}\\ 
    \midrule[1pt]
    \midrule[1pt]
    
    \multirow{5}{*}{MobileNetV3-Large} & - & - & 76.97 & 2.76 \\
    \cline{2-5}
    \multirow{5}{*}{} & \multirow{4}{*}{DUET} & - & 79.49 & 2.84\\
     \cline{3-5}
    \multirow{5}{*}{} & \multirow{4}{*} & 2 & 80.40 & 4.26 \\ \cline{3-5}
    \multirow{5}{*}{} & \multirow{4}{*} & 3 & 81.88 & 2.81 \\ \cline{3-5}
    \multirow{5}{*}{} & \multirow{4}{*} & 5 & \textbf{82.45} & \textbf{2.67} \\ 
     \midrule[1pt]
    \multirow{5}{*}{MobileNetV3-Small} & - & - & 67.88 & 2.15 \\ \cline{2-5}
    \multirow{5}{*}{} & \multirow{4}{*}{DUET} & - & 71.82 & 2.91 \\
     \cline{3-5}
    \multirow{5}{*}{} & \multirow{4}{*} & 2 & 73.11 & 3.80 \\ \cline{3-5}
    \multirow{5}{*}{} & \multirow{4}{*} & 3 & 73.74 & 2.19 \\ \cline{3-5}
    \multirow{5}{*}{} & \multirow{4}{*} & 5 & \textbf{75.15} & \textbf{1.41} \\ 
     \bottomrule[2pt]
    \end{tabular}
    }
    \label{tab:moe_num_affect_acc}
\end{table}

We also present the Param. (size of parameters) and FLOPs (floating-point operations per second) in Table~\ref{tab:experiment_rs}. Intuitively, models with low FLOPs and parameters are easier to deploy on devices. Note that the Parameters and FLOPs shown in the table are the primary models that need to be deployed on the device. Since PPG is deployed on the cloud, the parameters and FLOPs of the primary model under DUET framework are the same as the baselines.

\noindent \textbf{Results of Facial Expression Recognition.}
Table~\ref{tab:experiment_cv} reports the performance comparison between our model and the adopted baselines on the facial expression recognition task. The conclusions are generally consistent with the experiments on the recommendation task. The main differences are: (1) Since the annotation of real-time on-device samples is not routinely available, the fine-tuning experiments cannot be performed. (2) Compared with DUET (w/o SWA), the performance improvement of DUET on vision tasks is more significant than the recommendation task. Our intuition is that the real-time data distribution gap between real-time data in the computer vision task is smaller than the recommendation task, where the SWA module can further boost the parameters generation-based DMG framework.

Summing up, the results presented above confirm
the superiority of the proposed tuning-free device-cloud collaborative parameters
generation framework, which exhibits a faster and more accurate DMG paradigm simultaneously.

\subsection{In-Depth Analysis.}
We conducted the additional experiments on the \texttt{CK+} dataset to verify the strength of the proposed DUET.

\noindent \textbf{Detail Performance Analysis.}
To further study the effectiveness of DUET, we visualize the accuracy and loss in training and inference in Figure~\ref{fig:acc_loss_error_bar}, the corresponding mean value, median value, standard derivation, maximum/minimum, and quartiles of real-time inference are shown in Figure~\ref{fig:performance_boxplot}. All experiments are repeated five times. As shown in Figure~\ref{fig:acc_loss_error_bar}, the loss of DUET and DUET (w/o SWA) decreases faster than baseline, demonstrating its superior convergence speed in training. However,  this figure also indicates that the DUET (w/o SWA) brings a huge performance improvement but alone with the prediction oscillation problem. In addition, DUET with the SWA kindly solved this problem, resulting in a stable prediction in training and inference, which verifies the effectiveness of the proposed SWA module. More analysis about the stability is shown in sec.~\ref{sec:sta}. In Figure~\ref{fig:performance_boxplot}, the dark-colored line is the Mean value of the data, and the light-colored area is the fluctuation range, that is, the Maximum and Minimum values. For more clarity, the curves in the figure are obtained by 1:25 sampling, \emph{i.e.}, one epoch in the figure represents the actual 25 epochs. It also shows that DUET achieves consistent superiority in terms of all of the metrics.

\noindent \textbf{Generalizability.} Figure~\ref{fig:before_and_after_updating} depicts the performance change of different architectures over distribution shifts. Specifically, during two adjacent sessions (\textit{e.g.}, 1 \& 2), we simulate the distribution shifts by selecting heterogeneous data samples. At the beginning of each session, we update the base models with DUET and DUET (w/o SWA). MobileNetV3-Large (w/o DMG) refers to the base model trained on the cloud with all device data without any DMG methods. According to the results, we observe that DUET shows consistent performance improvement over the baseline against distribution shifts, which demonstrate the high generalization capability of the proposed DUET.

\noindent \textbf{Prediction Stability of Different $\mathcal{N}_p$.}
\label{sec:sta}
To build insights of stability on the SWA module,  we perform the ablation study that sets different numbers $\mathcal{N}_p$ of SWAs. Our experiments were repeated five times to observe its effect on accuracy and loss. On the one hand, as shown in Table~\ref{tab:moe_num_affect_acc}, 
the best performance is achieved with $\mathcal{N}_p$=5, both in accuracy and standard deviation. When the $\mathcal{N}_p$ is smaller, the standard deviation higher - especially the DUET (w/o SWA) - produces a higher performance fluctuation than baseline (w/o DMG), which indicates that the SWA module can improve the stability in prediction. On the other hand, Figure~\ref{fig:training_loss_curve_moe_num} systematically presents the explicit benefits of the $\mathcal{N}_p$ increase conditioned on two baselines. The figure shows that as the number of SWAs increases, the accuracy curve and loss curve are more stable, which also effectively accelerates the convergence speed and improves the performance in training. These results empirically verified the robustness of the SWA module, which provides a reliable solution that guarantees prediction stability.

\noindent \textbf{Time Delay Analysis}
Table~\ref{tab:delivery_delay} shows the size and time delay of \emph{uploading} real-time samples and \emph{downloading} dynamic layer model parameters on the device. Across all settings, this table suggests that our method produces a low time delay from 0.08s $\sim$ 1.06s, which seems acceptable for real-time applications. It is noteworthy that if we upload the embedding of real-time samples, it can generate a faster DMG and protect the user privacy of these samples. The observation above and analysis verify the effectiveness of DUET in implementing the real-time requirement for applications, thereby rendering the practicability for on-device learning. 

%% file: tex/5conclusion.tex
\vspace{-0.05cm}
\section{Conclusion}
\label{sec:conclusion}
\begin{sloppypar}
In this paper, we propose the
DUET for efficient device model generalization by generating adaptive device model parameters from the cloud without on-device training. Our method effectively learns a mapping function from real-time samples to device model parameters, which yields a low-time delay and better device-specific personalization.  Extensive experiments conducted on \texttt{Movielens-1M}, \texttt{Movielens-100k} and \texttt{CK+} show that DUET outperforms fine-tuning methods by a large margin in terms of accuracy and real-time performance, which validates the potential value in practical applications.
\end{sloppypar}

%% file: tex/acknowledgement.tex
\section*{ACKNOWLEDGEMENTS}
\begin{sloppypar}
This work was supported in part by National Key Research and Development Program of China (2022YFC3340900), National Natural Science Foundation of China (U20A20387, No. 62006207, No. 62037001), Project by Shanghai AI Laboratory (P22KS00111), Program of Zhejiang Province Science and Technology (2022C01044), the StarryNight Science Fund of Zhejiang University Shanghai Institute for Advanced Study (SN-ZJU-SIAS-0010), Fundamental Research Funds for the Central Universities (226-2022-00142, 226-2022-00051), the National Research Foundation, Singapore under its Emerging Areas Research Projects (EARP) Funding Initiative. 
\end{sloppypar}